\documentclass[11pt]{article}
\usepackage[utf8]{inputenc}
\usepackage[T1]{fontenc}
\usepackage[british]{babel}
\usepackage{fancyhdr}
\usepackage{times}
\usepackage{sidecap}
\usepackage{geometry}
\usepackage{floatrow}
\usepackage{float}
\usepackage{hyperref}
\usepackage{graphics,graphicx}
\usepackage{subcaption}
\usepackage{epsf,wrapfig}

\usepackage[misc]{ifsym}

\usepackage{natbib,bm}

\usepackage{paralist}
\let\olditem\item
\renewenvironment{thebibliography}[1]{%
  \section*{\refname}
  \let\par\relax
  \renewcommand{\item}[1][]{\olditem[\textbullet]}%
  \inparaenum}{\endinparaenum}

\newcounter{aff}

\DeclareGraphicsExtensions{.jpg,.pdf,.png,.eps}

\usepackage{caption}
\usepackage{xcolor}

\usepackage{tcolorbox}

\bibpunct{(}{)}{;}{a}{}{,}

\usepackage{soul}

\definecolor{darkcyan}{rgb}{0.0, 0.0, 0.7}

\definecolor{boxcolor}{RGB}{193, 229, 255}
\definecolor{figboxcolor}{RGB}{193, 229, 255}
\definecolor{boxcolorgray}{RGB}{227, 227, 227}
\definecolor{darkcyan}{rgb}{0.0, 0.0, 0.7}

\sethlcolor{boxcolorgray}

\def\gtrsim{\lower 2pt \hbox{$\, \buildrel {\scriptstyle >}\over
{\scriptstyle \sim}\,$}}
\def\lesssim{\lower 2pt \hbox{$\, \buildrel {\scriptstyle <}\over
{\scriptstyle \sim}\,$}}

\def\approxlt{\lower.2em\hbox{$\buildrel < \over \sim$}}
\def\approxgt{\lower.2em\hbox{$\buildrel > \over \sim$}}

\def\kms{{\rm km\ s^{-1}}}
\def\be{\begin{equation}}
\def\ee{\end{equation}}
\def\bea{\begin{eqnarray*}}
\def\eea{\end{eqnarray*}}

\def\jref@jnl#1{{\rm#1}}
\def\aj{\jref@jnl{AJ}}			
\def\araa{\jref@jnl{ARA\&A}}		
\def\apj{\jref@jnl{ApJ}}			
\def\apjl{\jref@jnl{ApJ}}		
\def\apjs{\jref@jnl{ApJS}}		
\def\ao{\jref@jnl{Appl.~Opt.}}		
\def\apss{\jref@jnl{Ap\&SS}}		
\def\aap{\jref@jnl{A\&A}}		
\def\aapr{\jref@jnl{A\&A~Rev.}}		
\def\aaps{\jref@jnl{A\&AS}}		
\def\azh{\jref@jnl{AZh}}			
\def\baas{\jref@jnl{BAAS}}
\def\icarus{\jref@jnl{Icarus}}	
\def\jrasc{\jref@jnl{JRASC}}		
\def\memras{\jref@jnl{MmRAS}}		
\def\mnras{\jref@jnl{MNRAS}}		
\def\pra{\jref@jnl{Phys.~Rev.~A}}	
\def\prb{\jref@jnl{Phys.~Rev.~B}}	
\def\prc{\jref@jnl{Phys.~Rev.~C}}	
\def\prd{\jref@jnl{Phys.~Rev.~D}}	
\def\pre{\jref@jnl{Phys.~Rev.~E}}	
\def\prl{\jref@jnl{Phys.~Rev.~Lett.}}	
\def\pasp{\jref@jnl{PASP}}		
\def\pasj{\jref@jnl{PASJ}}		
\def\qjras{\jref@jnl{QJRAS}}		
\def\skytel{\jref@jnl{S\&T}}		
\def\solphys{\jref@jnl{Sol.~Phys.}}	
\def\sovast{\jref@jnl{Soviet~Ast.}}	
\def\ssr{\jref@jnl{Space~Sci.~Rev.}}	
\def\zap{\jref@jnl{ZAp}}			
\def\nat{\jref@jnl{Nature}}		
\def\iaucirc{\jref@jnl{IAU~Circ.}}
\def\aplett{\jref@jnl{Astrophys.~Lett.}}
\def\apspr{\jref@jnl{Astrophys.~Space~Phys.~Res.}}
\def\bain{\jref@jnl{Bull.~Astron.~Inst.~Netherlands}}
\def\fcp{\jref@jnl{Fund.~Cosmic~Phys.}}
\def\gca{\jref@jnl{Geochim.~Cosmochim.~Acta}}
\def\grl{\jref@jnl{Geophys.~Res.~Lett.}}
\def\jcp{\jref@jnl{J.~Chem.~Phys.}}	
\def\jgr{\jref@jnl{J.~Geophys.~Res.}}	
\def\jqsrt{\jref@jnl{J.~Quant.~Spec.~Radiat.~Transf.}}
\def\memsai{\jref@jnl{Mem.~Soc.~Astron.~Italiana}}
\def\nphysa{\jref@jnl{Nucl.~Phys.~A}}
\def\physrep{\jref@jnl{Phys.~Rep.}}
\def\physscr{\jref@jnl{Phys.~Scr}}
\def\planss{\jref@jnl{Planet.~Space~Sci.}}
\def\procspie{\jref@jnl{Proc.~SPIE}}

\def\ion#1#2{#1$\;${\small\rm\@Roman{#2}}\relax}

\newcount\lecurrentfam
\def\LaTeX{\lecurrentfam=\the\fam \leavevmode L\raise.42ex
\hbox{$\fam\lecurrentfam\scriptstyle\kern-.3em A$}\kern-.15em\TeX}
\def\sizrpt{
(\fontname\the\font): em=\the\fontdimen6\font, ex=\the\fontdimen5\font
\typeout{
(\fontname\the\font): em=\the\fontdimen6\font, ex=\the\fontdimen5\font
}}
\definecolor{dark}{rgb}{ 0.34, 0.09, 0.49}

\newcommand{\Mjup}{M_{\rm Jup}}
\newcommand{\kmsec}{\mathrm{km\,s^{-1}}}
\newcommand{\msec}{\mathrm{m\,s^{-1}}}

\usepackage{fancyhdr,url}
\usepackage{transparent}
\usepackage{tablefootnote}
\usepackage{comment}
\usepackage{mathpazo}

\begin{document}

\noindent
\begin{center}
{
\Huge\bfseries Planet formation on solar system scales \\[0.5em]
\Large\it\bfseries How do gas giants form and shape their planetary system architectures?
}
\end{center}

\noindent
F. Zagaria$^1$, A. F. Izquierdo$^2$, J. Bae$^2$, P. Griveaud$^1$, R. A. Booth$^3$, P. Curone$^4$, A. Johansen$^{5}$, S. Facchini$^6$, M. Benisty$^1$, T. Birnstiel$^7$, A. Booth$^8$, O. Chrenko$9$, I. Czekala$^{10}$, K. Doi$^{1}$, G. Guidi$^{11}$, M. Kama$^{12}$, Q. Kral$^{13}$, C. Law$^{14}$, G. Lodato$^6$, P. Luppe$^{15}$, E. Macias$^{16}$, L. Majumdar$^{17,18}$, R. Nealon$^{19}$, A. Penzlin$^7$, D. Polychroni$^{20}$, D. J. Price$^{19}$, Á. Ribas$^{21}$, R. Teague$^{22}$, C. Toci$^{23}$, D. Turrini$^{20}$, A. A. Sefilian$^{24}$, Y. Shibaike$^{25}$, D. Stamatellos$^{26}$, E. Szuszkiewicz$^{27}$, C. Walsh$^3$

\vspace{1em}
{
\footnotesize{
\noindent 
$^1$Max-Planck Institute for Astronomy (MPIA), Königstuhl 17, 69117 Heidelberg, Germany (\Letter\ \url{frzagaria@mpia.de});
$^2$Department of Astronomy, University of Florida, Gainesville, FL 32611, USA (\Letter\ \url{andres.izquierdo.c@gmail.com});
$^3$School of Physics and Astronomy, University of Leeds, Leeds, LS2 9JT, UK;
$^4$Departamento de Astronomía, Universidad de Chile, Camino El Observatorio 1515, Las Condes, Santiago, Chile;
$^5$Center for Star and Planet Formation, Globe Institute, University of Copenhagen, Øster Voldgade 5-7, 1350 Copenhagen, Denmark; 
$^6$Dipartimento di Fisica, Università degli Studi di Milano, Via Celoria 16, 20133 Milano, Italy;
$^7$University Observatory, Faculty of Physics, Ludwig-Maximilians-Universität München, Scheinerstr. 1, 81679 Munich, Germany;
$^8$ Center for Astrophysics Harvard \& Smithsonian, 60 Garden Street, Cambridge, MA 02138, USA;
$^9$Charles University, Faculty of Mathematics and Physics, Astronomical Institute, V Holešovickách 747/2, 180 00, Prague 8, Czech Republic;
$^{10}$Australia Telescope National Facility, CSIRO Space and Astronomy, Australia;
$^{11}$Univ. Grenoble Alpes, CNRS, IPAG, F-38000 Grenoble, France;
$^{12}$Department of Physics and Astronomy, University College London, Gower Street, London WC1E 6BT, UK;
$^{13}$LIRA, Observatoire de Paris, Université PSL, Sorbonne Université, Université Paris Cité, CY Cergy Paris Université, CNRS, France;
$^{14}$Minnesota Institute for Astrophysics, University of Minnesota, US;
$^{15}$School of Physics, Trinity College Dublin, the University of Dublin, College Green, Dublin 2, Ireland;
$^{16}$European Southern Observatory (ESO), Karl-Schwarzschild-Straße 2, Garching bei München, Germany;
$^{17}$National Institute of Science Education and Research, Odisha, India;
$^{18}$Homi Bhabha National Institute, Mumbai, India;
$^{19}$School of Physics and Astronomy, Monash University, Clayton VIC 3800, Australia;
$^{20}$INAF - Turin Astrophysical Observatory, IT;
$^{21}$School of Physical and Chemical Sciences, Queen Mary University of London, Mile End, London E1 4NS, UK;
$^{22}$Department of Earth, Atmospheric and Planetary Sciences, Massachusetts Institute of Technology, Cambridge, MA 02139, USA;
$^{23}$Departamento de Física Aplicada III, Universidad de Sevilla, Camino descubrimientos s/n, 41092, Sevilla, Spain;
$^{24}$Department of Astronomy and Steward Observatory, University of Arizona, Tucson, AZ 85721, USA;
$^{25}$Graduate School of Science and Engineering, Kagoshima University, JP; 
$^{26}$Institute for Mathematics, Physics \& Astronomy, University of Lancashire, Preston, PR1 2HE, UK;
$^{27}$Institute of Physics and CASA*, University of Szczecin Wielkopolska 15, PL-70-451, Szczecin, Poland.
}
}

\begin{tcolorbox}[boxsep=3pt,left=0pt,right=0pt,top=1pt,bottom=1pt,colframe=boxcolor,colback=boxcolor]
\justify
\textbf{Abstract:} 
Giant planets are often regarded as primordial protoplanetary ``architects''. Their formation and interactions with planetesimals and other planets during and after the protoplanetary disc phase are key to setting the blueprint of the planetary systems we observe today.~Catching nascent gas giants is our best opportunity to uncover these crucial steps in the planet formation process and shed light on the origins of our own solar system. Over the last decade, ALMA has revolutionised our understanding of planet formation; still, detecting signatures~of~Jupiter-like planets has remained out of reach. We argue that ALMA2040, a next-generation sub(mm) interferometer with improved sensitivity ($\times25$ in the continuum and $\!\times10$ for line emission)~and~a~$\!\times3$ higher angular resolution compared to ALMA, will be sensitive to (sub)mm emission~around~a forming~Jupiter-like~planet orbiting a Sun-like star within $150\,{\rm pc}$ and allow kinematically characterising interactions of Jovian and sub-Jovian planets with their natal disc. ALMA2040 will provide unique and unprecedented insights into how gas giants carve gaps, accrete material, form satellites, and affect the architectures of their co-evolving planetary systems.
\end{tcolorbox}

\newpage{}

\section{Scientific context and current challenges}\label{sec:intro}\vspace{-5.0pt}
The several thousand exoplanets detected over the past decades display a remarkable diversity \citep{Zhu&Dong2021} thought to be a consequence of their formation. Cold gas giants play a key~role in planet formation: these planets have been detected in many systems~and suggested to actively shape their formation environments. Jupiter is a compelling example: its interactions with other planets and leftover planetesimals in the protosolar disc are routinely invoked to explain the properties and architecture of the present-day solar~system \citep[e.g.,][]{Turrini2011, Raymond&Izidoro2017,Pirani2019}. Catching cold gas giants in their infancy is thus our best opportunity to understand their impact on planetary system formation, including our own solar system.

Simulations show that forming gas giants leave clear imprints on their surroundings. They~excite spirals and open up annular gaps where drifting pebbles pile up in bright rings, and accrete material, potentially mediated by circumplanetary discs (CPDs),~through gap-crossing or meridional flows \citep[e.g.,][]{Bae2023}.$\,$These hallmarks~can~be~used~to~detect~(emission from) embedded proto-Jupiters and characterise their interactions with their natal disc.

ALMA has revolutionised our understanding of protoplanetary discs and planet formation~like no other facility. One of its most significant breakthroughs was the discovery that planet-forming discs ubiquitously display substructures, most commonly gaps and rings \citep[e.g.,][]{ALMAPartnership2015, Long2018, Andrews2018}. Embedded protoplanets have been proposed to be a natural explanation for these substructures; still detecting such planets proved challenging. Except for PDS~70~\citep{Keppler2018,Haffert2019} and WISPIT~2 \citep{vanCapelleveen2025,Close2025,Lawlor2026}, the high extinction and contamination~from~the~disc~(e.g., \citealt{Cugno2025}) have hindered any detections in the IR. Similarly, (sub)mm~continuum emission co-located with embedded planets proved too faint and difficult to disentangle~from~their hosting disc \citep[e.g.,][]{Andrews2021} except for PDS~70c \citep{Isella2019, Benisty2021}.

These challenges motivated alternative approaches, first and foremost kinematic studies~of~the perturbations induced by planets in their natal disc's velocity field (\citealt{Teague2018,Pinte2018}, see review by \citealt{Pinte2023}). However, kinematics also struggled to detect planets because of the current spectral line sensitivity limitations and the degree of confusion induced by~the richness of detected structures, and, as of today, few kinematic planetary candidates have been identified \citep[e.g.,][]{Pinte2018,Izquierdo2022,Izquierdo2026_exoALMA}.

In the coming years, the advent of 30-m class telescopes will push the frontier of protoplanet~detection in the optical and IR to lower masses and smaller separations. A next-generation (sub)mm interferometer up to these standards will be essential to develop a comprehensive picture of how these planets accrete material, form their moons, and give shape to their hosting systems. In this short paper we discuss the requirements needed for a next-generation (sub)mm interferometer to detect continuum emission around a proto-Jupiter orbiting a nearby ($\approx150\,{\rm pc}$) Sun-like star (Sect.~\ref{sec:CPDs}) and kinematically characterise interactions of (sub-)Jovian planets with their natal discs (Sect.~\ref{sec:kinematics}). Finally, we touch upon possible synergies with current and future facilities (Sect.~\ref{sec:synergies}).\vspace{-10.0pt}

\section{Circumplanetary emission around a proto-Jupiter}\label{sec:CPDs}\vspace{-5.0pt}
Giant protoplanets grow by accreting material from their forming disc through an envelope~or~a~rotationally supported CPD. Thanks to the lower degree of contamination from the star and the~circumstellar material, (sub)mm observations, especially across a wide spectral range, provide unique insights into these systems, opening up the possibility of estimating the CPD mass budget to form satellites and better understanding how material is delivered to and accreted by protoplanets.

\begin{SCfigure}
    \includegraphics[width=0.65\textwidth]{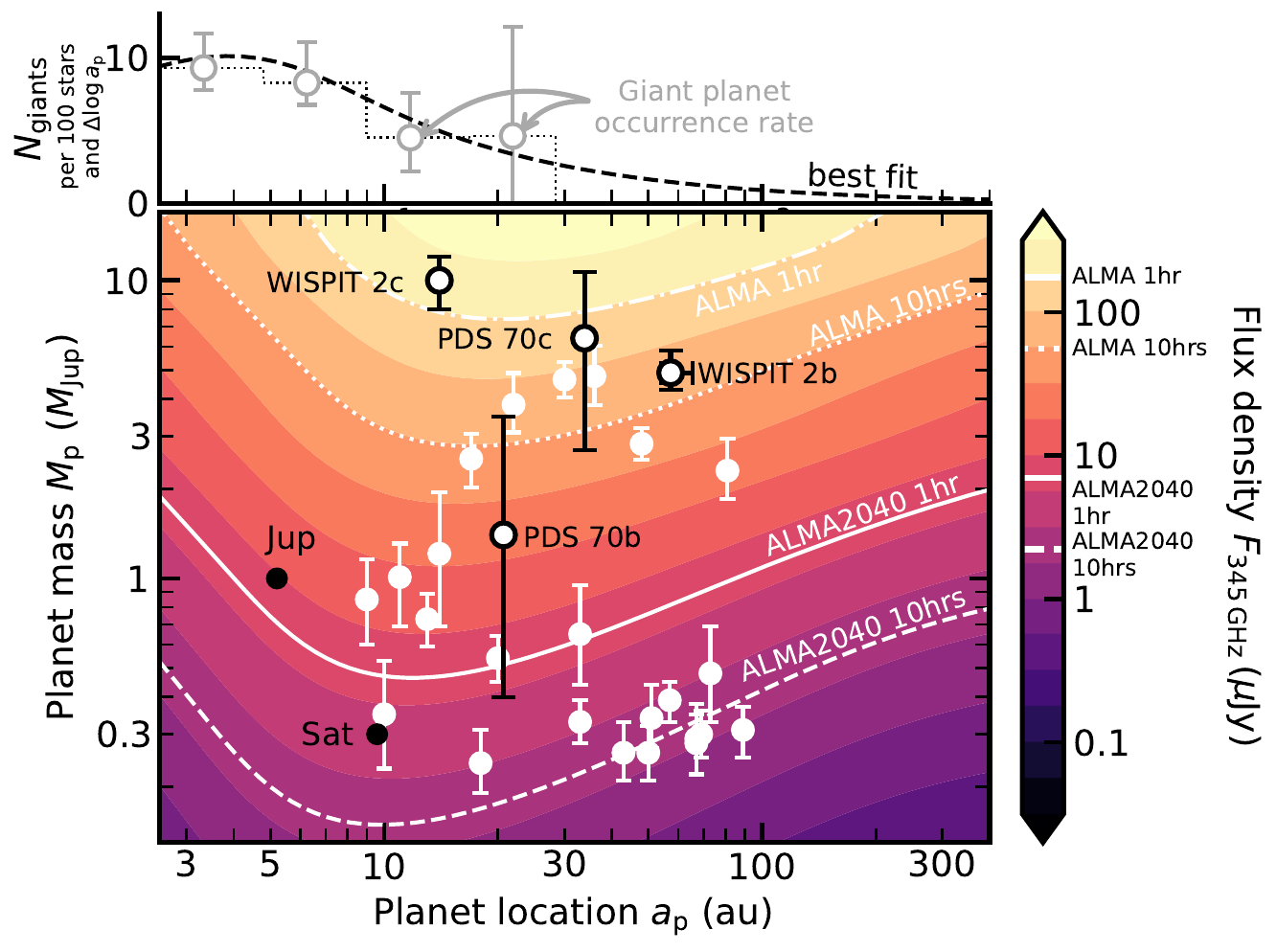}
    \caption{CPD continuum~flux densities$\;$at$\;345\,{\rm GHz}\;$for$\;$diffe-rent$\;$planet$\;$masses$\;$and$\;$separa-tions. The dots mark the solar system$\;$gas$\;$giants,$\;$the$\;$directly imaged$\;$planets$\;$in$\;$PDS$\;$70$\;$and WISPIT$\;$2,$\;$and$\;$candidate$\;$proto-planets$\;$in disc gaps. For$\;$com-parison,$\;$the$\;$distribution$\;$of$\;$ful-ly formed gas giants is shown (top panel). The white curves display sensitivity limits.}
    \label{fig:CPDmm}
    \vspace{-10.0pt}
\end{SCfigure}

We explored the detectability of CPDs around giant planets orbiting a nearby ($150\,{\rm pc}$) Sun-like star over a range of masses and separations. To give quantitative estimates of their (sub)mm~flux density, we adopted the semi-analytical models of \citet{Andrews2021}. In these models,~the~CPD temperature is set by a combination of stellar/planet irradiation and planetary accretion heating. These contributions were estimated by adopting predictions for planet luminosities and radii from evolution models \citep{Spiegel&Burrows2012, Linder2019} and simple scaling relations. The CPD extends to a third of the planet's Hill's sphere radius and its density scales as a power-law normalised to a fraction of the planet mass set to be consistent with the mm flux density of PDS~70c for grains of $1\,{\rm mm}$ and DSHARP opacities \citep{Birnstiel2018}. The colour map in Fig.~\ref{fig:CPDmm} shows the predicted CPD continuum flux densities at $345\,{\rm GHz}$. The shape of the isocontours is set by the combination of CPD temperature and size. For a fixed planet mass, CPD flux densities increase with planet separation as the CPD size also increases until they become colder further from the star. Where this transition takes place depends on the relative contribution of stellar and planetary heating, with the latter being more relevant for more massive planets.

Overlaid on Fig.~\ref{fig:CPDmm} are the the solar system gas giants, the directly imaged planets in PDS~70 and WISPIT~2 \citep{Trevascus2025, vanCapelleveen2025, Lawlor2026}, and the protoplanets proposed to be carving the gaps detected in high angular resolution ALMA continuum observations \citep{Ruzza2025}. The plot shows that ALMA is sensitive to a planet population fundamentally different from the bulk of mature gas giants ($M_{\rm p}\sin i\approx0.1-20\,M_{\rm Jup}$) detected to date, whose~occurrence rate peaks around $4\,{\rm au}$ \citep[][top panel of Fig.~\ref{fig:CPDmm}]{Fulton2021}.

The white tracks in Fig.~\ref{fig:CPDmm} illustrate $7\sigma$-significance detection limits. The dashed-dotted and~dotted lines correspond to ALMA on source integrations of $1\,{\rm hr}$ (similar to the DSHARP ALMA Large Program, \citealt{Andrews2018}) and $10\,{\rm hrs}$ (for illustrative purposes). These tracks indicate that detecting circumplanetary emission is only feasible around the most massive planets. Extending~the sample of detectable CPDs to the bulk of gas giants ($M_{\rm p}\lesssim M_{\rm Jup}$, $a_{\rm p}\!\!<\!\!10\,{\rm au}$) requires a combination of $\times3$ higher angular resolution and $\times25$ point-source sensitivity. The solid and dashed lines display $\times25$ deeper detection limits compared to ALMA. These tracks highlight the possibility of detecting (sub)mm continuum emission around a Jupiter-like planet in moderate integrations ($1\,{\rm hr}$ on source), and around a Saturn-mass planet for extremely deep observations ($10\,{\rm hrs}$ on source), corresponding to a moon-forming budget of $\approx50$ and $15$ Ganymede masses. \vspace{5.0pt}

\noindent \textbf{Point-source injection/recovery experiments:} Fig.~\ref{fig:CPDmm} shows that a continuum sensitivity $\times25$ higher than ALMA could allow detecting (sub-)mm continuum emission around a proto-Jupiter. However, detecting CPDs is challenging not only because of their faint (sub-)mm emission, but also because it requires disentangling circumplanetary and circumstellar disc emission \citep{Andrews2021}. This can be difficult especially closer to the star, where planetary gaps are difficult to resolve even for high-angular resolution observations. To better assess the impact of these effects,~we carried out point source (PS) injection-recovery experiments.

We ran 2D hydrodynamic simulations of the interactions of a $1M_{\rm Jup}$ planet orbiting a Sun-like star with its hosting disc using FARGO3D \citep{Benitez-Llambay&Masset2016}. Simulations~span a range of disc properties affecting the shape and contrast of planetary-driven substructures, including aspect ratio and viscosity, with a $\beta=1$ cooling prescription. They include 13 dust species with constant grain sizes logarithmically spaced between $8\times10^{-3}$ and $8\,{\rm cm}$. The simulation outputs were post-processed using the Monte Carlo radiative transfer code \texttt{RADMC-3D} \citep{Dullemond2012}. For each parameter combination, we first computed the disc thermal structure for different planet locations (taking care of scaling densities and grain sizes in such a way to preserve the Stokes number at the reference radius) and then ray-traced images at the reference frequency of $345\,{\rm GHz}$. Mock observations were generated for these sky models using \texttt{simobserve} \citep{CASATeam2022} for 
$1\,{\rm hr}$ on source and a PWV of $1\,{\rm mm}$, for an example ALMA2040 array of 255 $12\,{\rm m}$ antennas in an extended and intermediate configuration over a $128\,{\rm GHz}$ bandwidth.

\begin{figure*}[b!]
    \vspace{-10.0pt}
    \centering
    \includegraphics[width=\linewidth]{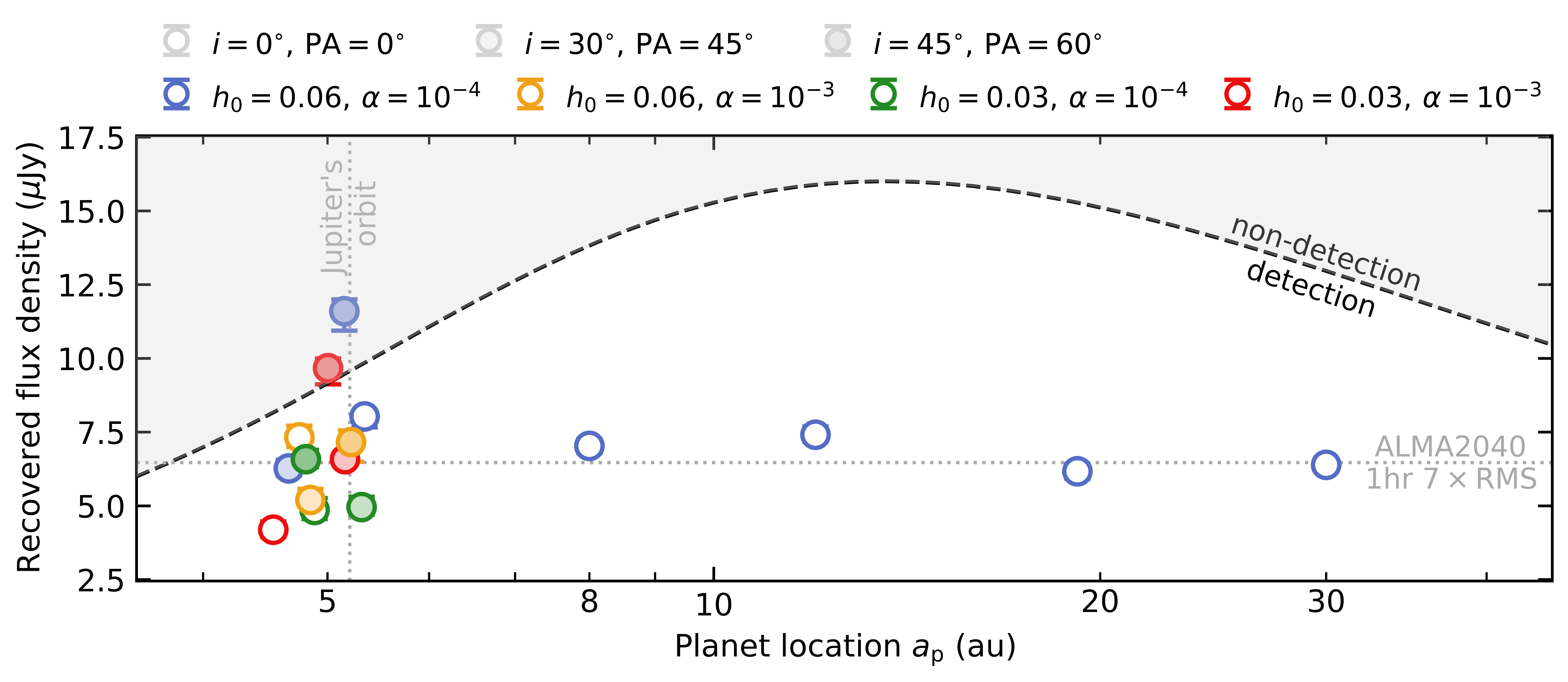}
    \caption{Minimum recovered PS flux for different planet positions, disc geometries, and simulation parameters. The black dashed line displays the predicted flux densities of a $1\,M_{\rm Jup}$ planet.}
    \label{fig:inj-rec}
\end{figure*}

These datasets worked as the basis for injection-recovery experiments for which we adopted the pipeline published by \citet{Andrews2021}. 100 point sources of prescribed flux density were injected in the simulated visibilities randomly in azimuth and radius within the planetary gap for a range of luminosities. The resulting visibilities were then modelled non-parametrically to subtract the axisymmetric disc component using \texttt{frank} \citep{Jennings2020}. We searched for the injected point sources in the residuals imaged with \texttt{tclean}. 
Fig.~\ref{fig:inj-rec} displays the minimum injected PS flux density for which we measured a recovery fraction of 50\% for different planet positions, disc geometries, and simulation parameters.

For $a_{\rm p}>8\,{\rm au}$ the planetary gap is resolved by at least $50$ independent beams and the PS detection thresholds cluster around $7\times$ the predicted noise level (same as the solid line in Fig.~\ref{fig:CPDmm}). Instead, for $a_{\rm p}=5\,{\rm au}$ high-significance residuals due to lingering disc emission and planetary-induced spirals increase the CPD detection threshold, especially for higher inclinations.~These~effects are sensitive to the shape and contrast of planet-induced gap and spirals, as indicated by the dependence of the PS recovery threshold on viscosity and aspect ratio. Fig.~\ref{fig:inj-rec} also displays the flux density~of a CPD around a Jupiter-mass planet predicted by the models introduced in the previous section (dashed black line). These flux densities are well above recovery fractions of 50\% except for models with the highest inclinations at $5\,{\rm au}$. This indicates that a next-generation interferometer with $\times3$ higher angular resolution and $\times25$ point-source sensitivity than ALMA could be sensitive to circumplanetary emission around close-by Jupiter-like planets.\vspace{5.0pt}

\noindent \textbf{Caveats:} We generated mock observations over an instantaneous bandwidth of $128\,{\rm GHz}$. Our efforts to also consider the frequency dependence of circumplanetary and circumstellar disc emission over this wide spectral range proved prohibitively time consuming and are not discussed, but will be essential in future more thorough explorations. The goal for ALMA2040 is to extend the instantaneous bandwidth to over $512\,{\rm GHz}$. Assessing the effects of such a wide spectral coverage on CPD detection is not straightforward. Increasing continuum sensitivity will also make detecting asymmetric planetary-driven features easier. While this could help identify the planet position and restrict the CPD search area, their contamination might not necessarily result in lower CPD detection thresholds. Finally, we caveat that our results are dependent on the range of validity of the assumed CPD models. For example, they predict that it should be relatively easy to detect circumplanetary emission around WISPIT~2c, which instead we do not see \citep{Benisty2026}. This is suggesting that other factors, such as details of the planetary accretion process, the CPD evolution timescales, and replenishment from the circumstellar disc likely play a key role in CPD detectability. The sensitivity and resolution improvements we discussed will be essential to test these assumption, build better CPD models, and understand material delivery to protoplanets.\vspace{-10.0pt}

\section{Kinematic detection of embedded planets with ALMA2040}\label{sec:kinematics}\vspace{-5.0pt}
Young planets are difficult to detect directly because their emission is strongly attenuated by the surrounding circumstellar material. However, there is another way to pinpoint these elusive objects. The dynamical response of the disc provides a complementary route that is independent of the intrinsic luminosity of the planet. An embedded planet perturbs the local velocity field, launches spiral wakes, modifies the pressure-supported rotation across its gap, and drives three-dimensional flows in its vicinity \citep[see][for a review]{Pinte2023}. When observed in spectrally resolved molecular-line emission, these motions produce spatially and spectrally coherent deviations from an otherwise smooth Keplerian disc, including localised velocity perturbations, excess line broadening, and asymmetric line profiles \citep[e.g.,][]{Izquierdo2026_exoALMA}. Recent ALMA observations of WISPIT~2 reveal gas gaps and localised
kinematic signatures associated with the directly imaged planet
WISPIT~2b, with an estimated mass of $\approx\!5\,\Mjup$
\citep{vandermarel+2026,Benisty2026}, providing an observational
benchmark linking embedded planets to the dynamical response
of their host discs. Our sensitivity forecast suggests that ALMA2040 could extend molecular-line kinematic searches from Jupiter-mass planets into the sub-Jovian regime within a few hours of integration.\vspace{-10.0pt}

\subsection{Resolving planet-driven flows}\label{sec:planetflows}\vspace{-5.0pt}
To quantify this advance, we compare synthetic ALMA2040 observations with an ALMA reference image generated under fiducial exoALMA observing conditions. The exoALMA Large Program targets the dynamical structure of discs in CO at approximately $0.15''$ angular resolution, with $0.1\,\kmsec$ channels and RMS sensitivities of order $1$--$2\,$K per channel \citep{teague+2025}. At $150\,$pc, however, one resolution element still corresponds to roughly $20$--$25\,$au. The flows associated with a planet, its spiral wakes, the pressure gradient across the gap, and emission arising at different disc heights are thus mixed within the same beam, which often leads to substantial modulation of the planet-driven signal and biases current kinematic detections towards Jupiter-mass planets \citep{Izquierdo2026_exoALMA}. Resolved observations of multiple molecular species and transitions with ALMA2040 could probe these flows at different disc heights and help distinguish planet-driven three-dimensional motions from apparent kinematic perturbations associated with surface geometry or thermochemical structure. 

Although smaller beams can be achieved with ALMA, obtaining the line sensitivity required for precision kinematics generally forces observations towards more moderate angular resolution. This compromise has two important consequences. First, present observations primarily reveal the large-scale response of the disc rather than the flow in the immediate vicinity of the planet. Beam convolution dilutes localised perturbations and complicates their separation from emission-surface variations, warps, disc instabilities, or other non-planetary structures. Second, current kinematic searches are biased towards planets on wide orbits, where their signatures are more readily separated from the central disc; at smaller radii, limited angular resolution exacerbates velocity mixing. This leaves the inner tens of au (where giant planets are expected to undergo much of their growth and migration) largely unexplored. \vspace{5.0pt} 

\noindent\textbf{Reference synthetic observations.} ALMA2040 will break this resolution-sensitivity compromise. Our reference calculation adopts a planet-disc interaction simulation with a planet-to-star mass ratio $M_{\rm p}/M_*=5\times10^{-4}$, corresponding to approximately $1\,\Mjup$ around a $2\,M_\odot$ star \citep[see][for the setup]{Izquierdo2026_exoALMA}. Starting from a noiseless radiative-transfer cube of the $^{12}$CO $J=2$--$1$ line for a near face-on disc at $i=15^\circ$ and $d=101\,$pc, we generated ALMA2040 visibilities with CASA \texttt{simobserve} combining compact and intermediate configurations. The seven-hour on-source integration was divided in a 20:80 ratio between the two configurations. We imaged the combined visibilities with multiscale \texttt{tclean}, using a channel-dependent emission mask, Briggs robust parameters of $+0.5$ and $0.0$, and a stopping threshold of three times the per-channel RMS. 

The resulting Briggs $+0.5$ and Briggs $0.0$ images reach beams of $0.095''\!\times\!0.065''$ and $0.052''\!\times\!0.046''$, respectively. In $0.1\,\kms$ channels, their reference seven-hour RMS values are $0.24\,$mJy beam$^{-1}$ ($0.9\,$K) and $0.27\,$mJy beam$^{-1}$ ($2.5\,$K). The matched exoALMA-like reference has a $0.15''\!\times\!0.15''$ beam and an RMS of $2.34\,$mJy beam$^{-1}$ ($2.4\,$K) per channel. Briggs $+0.5$ provides a good compromise between line sensitivity and angular resolution, whereas Briggs $0.0$ more sharply resolves the spatial morphology of the velocity perturbations.
The matched exoALMA reference and the controlled integration-time experiments were produced with \textsc{syndisk}\footnote{\url{https://github.com/richteague/syndisk}}, retaining the same sky model while applying the prescribed beam and injected thermal noise. We repeat the velocity-precision and planet-mass analysis at $i=30^\circ$, using the same beam, channel spacing, and noise. \vspace{5.0pt} 

\begin{figure}[t]
  \centering
  \includegraphics[width=\linewidth]{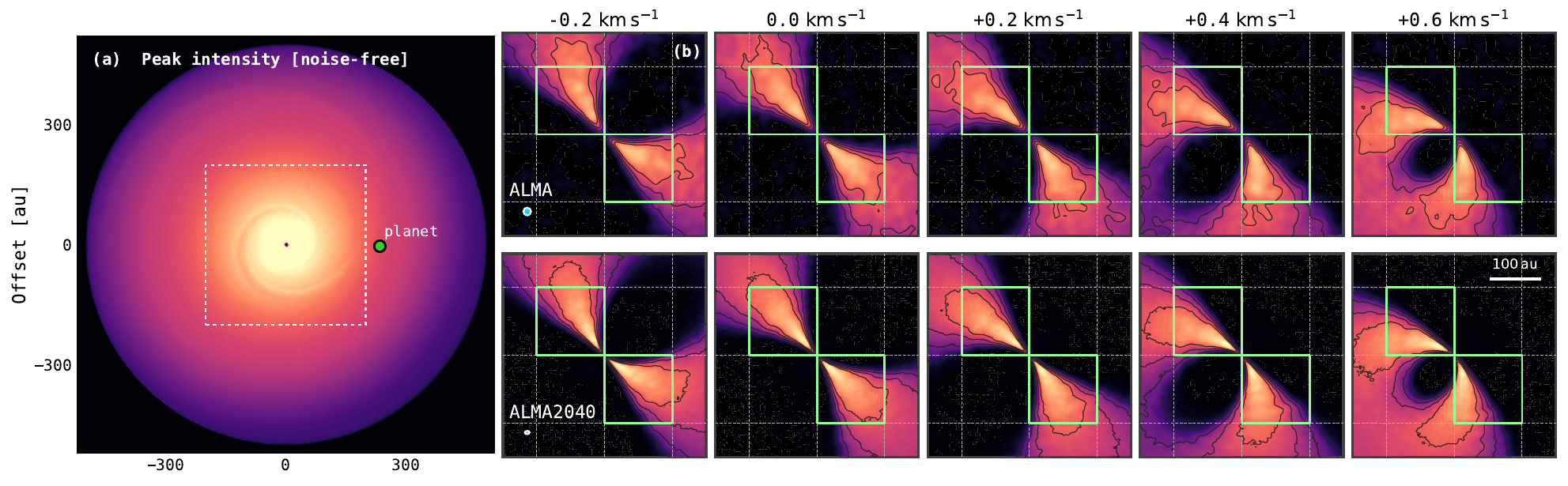}
  \caption{Planet-driven perturbations manifest as ``kinks'' in intensity channels, as illustrated for the $i=15^\circ$ reference model. \textit{(a)} Peak intensity of the noiseless, unconvolved radiative-transfer cube. The dashed square denotes the $\pm200\,$au region displayed in the channel panels, while the green marker indicates the projected location of the $1\,\Mjup$ planet at $R_{\rm p}=240\,$au. \textit{(b)} Intensity channels, plotted every $0.2\,\kms$, from the seven-hour exoALMA-like observation (top) and ALMA2040 Briggs $+0.5$ observation (bottom). Black contours are drawn at $19$, $34$, $49$, and $64\,$K. The soft-green squares identify the two central diagonal regions where the kink morphology is most apparent. The smaller beam and higher SNR of the ALMA2040 image preserve a sharper displacement of the emission contours and reveal its coherence across the sampled channels.}
  \label{fig:kinematic_kinks}
  \vspace{-10.0pt}
\end{figure}

\noindent\textbf{Recovering planet-driven structure.} As illustrated in Fig.~\ref{fig:kinematic_kinks}, the improved angular resolution and sensitivity of ALMA2040 make the planet-driven intensity wiggles, or ``kinks'', sharper and reveal their coherence across adjacent velocity channels. Although some planet-driven kinks are also visible in the ALMA channels, they are difficult to distinguish from noise-induced wiggles elsewhere in the disc and would therefore likely be characterised as non-detections. 

\begin{figure}[t]
  \centering
  \includegraphics[width=\linewidth]{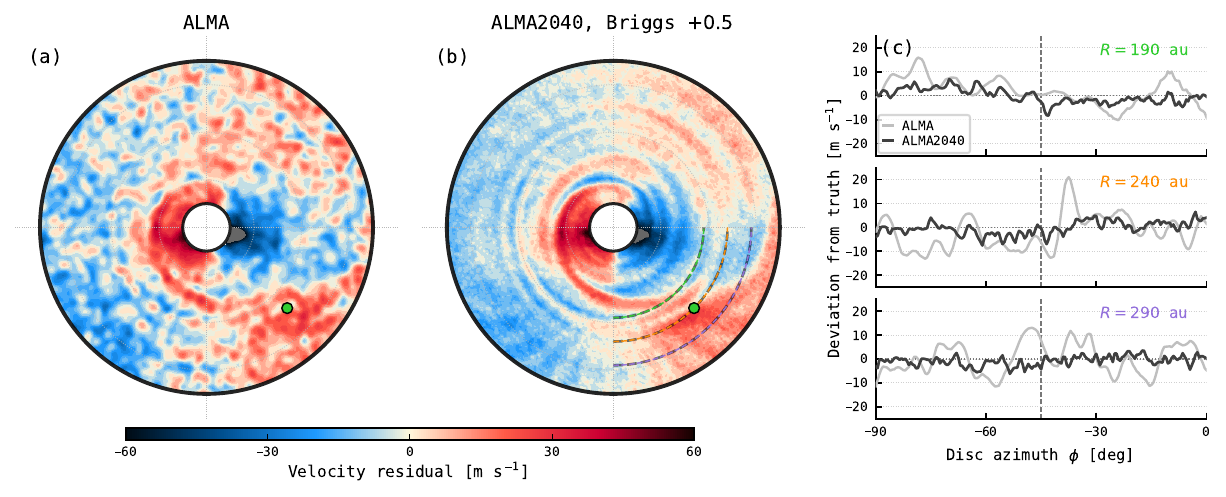}
  \caption{Centroid-velocity residuals and recovery of the planet-driven perturbations for the $i=15^\circ$ reference model. \textit{(a)} ALMA and \textit{(b)} ALMA2040 Briggs $+0.5$ residual maps after subtracting a smooth \textsc{discminer} Keplerian model, shown in deprojected disc-plane coordinates with a common $\pm60\,\msec$ colour scale. The green marker denotes the planet at $(R_{\rm p},\phi_{\rm p})=(240\,{\rm au},-45^\circ)$. The coloured dashed arcs in panel \textit{(b)} mark the azimuthal profiles extracted at $R=190$, $240$, and $290\,$au over $-90^\circ\leq\phi\leq0^\circ$. \textit{(c)} Deviation of the ALMA (light grey) and ALMA2040 (dark grey) profiles from the observational truth along the same arcs. Here, ``truth'' is the fitted centroid residual from the noiseless, unconvolved radiative-transfer cube and therefore includes line-of-sight projection, radiative-transfer weighting, and spectral sampling. 
  The vertical dashed line marks the azimuthal location of the planet. The smaller deviations from zero demonstrate that ALMA2040 more closely recovers both the amplitude and morphology of the planet-driven perturbation.}
  \label{fig:kinematic_residuals}
  \vspace{-10.0pt}
\end{figure}

In Fig.~\ref{fig:kinematic_residuals}, we compare centroid-velocity maps from the synthetic observations with those from a smooth \textsc{discminer}\footnote{\url{https://github.com/andizq/discminer}} Keplerian model to quantify the residuals contributing to the channel-map kinks. Residuals from the ALMA2040 image trace both the local Doppler flip triggered by the planet's wakes and the more extended response of the spiral structures away from the planet. In the ALMA image, only the strongest spiral perturbation (interior to the planet's orbit) is clearly recovered, but the remaining filamentary structures are confused by background fluctuations. Azimuthal slices through the wake at three radii show how the ALMA2040 image preserves both the amplitude and morphology of the velocity perturbations. 

\subsection{From centroid velocity precision to planet-mass sensitivity}\vspace{-5.0pt}
The gain in line sensitivity can be translated into the precision with which local gas velocities are measured. We use the seven-hour ALMA2040 Briggs $+0.5$ cube as the reference and scale the injected RMS to other integration times following
\begin{equation}
\sigma_I(t)=\sigma_I(7\,\mathrm{h})\left(\frac{7\,\mathrm{h}}{t}\right)^{1/2}.
\end{equation}
We adopt integration times $1$, $3$, $7$, $14$, and $28\,$h and generate ten independent noise realisations at each time with \textsc{syndisk}. The same noiseless sky model and beam are retained throughout, so the experiment isolates thermal sensitivity. 
At $i=15^\circ$, the median pixel-by-pixel scatter among the recovered Gaussian velocity maps follows a $t^{-1/2}$ scaling, decreasing from approximately $6.0\,\msec$ at $1\,$h to $3.5$, $2.3$, $1.6$, and $1.1\,\msec$ at $3$, $7$, $14$, and $28\,$h, respectively. At seven hours, the corresponding ALMA precision is $6.4\,\msec$, nearly three times poorer than ALMA2040 Briggs $+0.5$. These values are measured across the ten noise realisations in the planet-wake region. The equivalent off-planet control differs by less than $3\%$ at every integration time, confirming that this statistic, $\sigma_{v,\rm fit}$, traces the random centroid velocity precision rather than the local wake morphology. \vspace{5.0pt} 

\noindent\textbf{Planet-mass sensitivity.} To relate this precision to planet mass, we assume that the beam-convolved velocity peak scales linearly with planet-to-star mass ratio, $q=M_{\rm p}/M_*$, and calibrate its normalisation directly from the reference simulation:
\begin{equation}
v_{\rm pk}^{\rm beam}(q)\simeq
v_{\rm pk,0}^{\rm beam}\frac{q}{q_0},
\label{eq:vpeak_mass}
\end{equation}
where $q_0=5\times10^{-4}$ and the ALMA2040 Briggs $+0.5$ beam gives $v_{\rm pk,0}^{\rm beam}=29.7\,\msec$ at $i=15^\circ$ and $46.1\,\msec$ at $i=30^\circ$. These normalisations already incorporate line-of-sight projection, radiative-transfer weighting, and beam attenuation; only the linear dependence on $q$ is assumed, motivated by the low-mass scaling of \citet{rabago+2021}. 
We note that this relationship is specific to the adopted viewing geometry. If vertical motions are subdominant in this low-mass regime, the perturbation projects approximately as $\delta v_{\rm los}\simeq\sin i\left(\delta v_\phi\cos\phi+\delta v_R\sin\phi\right)$, so the inclination and planet azimuth change the relative contributions of the radial and azimuthal wake motions. In the native projected velocities of \citet{izquierdo+2021}, the largest peaks occur generally at an azimuthal planet location near $\phi_{\rm p}=0^\circ$, where $\delta v_\phi$ dominates. We therefore calibrate Equation~\ref{eq:vpeak_mass} separately at $i=15^\circ$ and $30^\circ$, both at $\phi_{\rm p}=-45^\circ$. 

\begin{figure}[t]
  \centering
  \includegraphics[width=\linewidth]{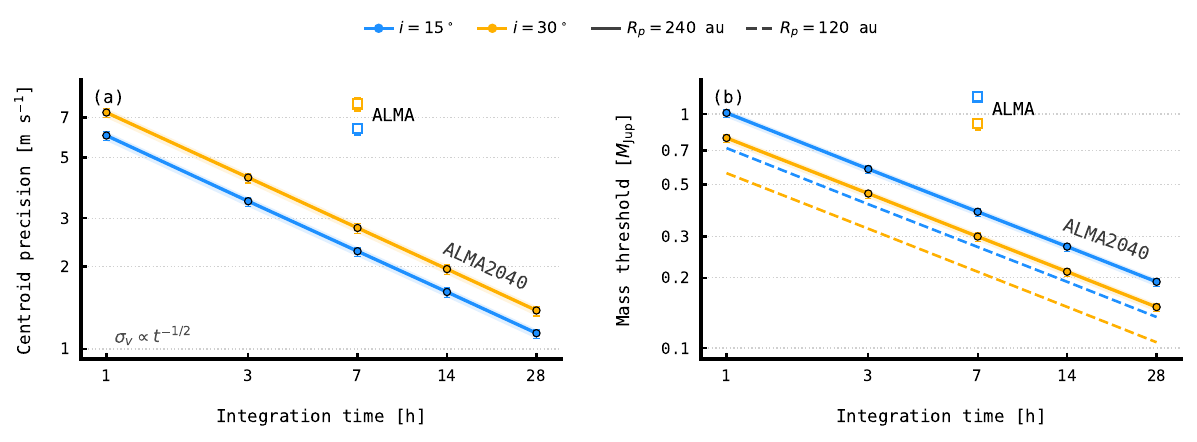}
  \caption{ALMA2040 centroid precision and kinematic planet-mass sensitivity at disc inclinations $i=15^\circ$ (blue) and $30^\circ$ (gold). \textit{(a)} Median centroid scatter across ten noise realisations; bands show 16th-84th percentile bootstrap uncertainty. Open squares denote the ALMA references. \textit{(b)} $5\sigma$ thermal-noise mass thresholds for a $2\,M_\odot$ host, assuming linear mass scaling of the full beam-convolved centroid peak. Solid curves adopt a planet orbital radius of $R_{\rm p}=240\,$au; dashed curves extrapolate to $120\,$au using $v_{\rm K}\propto R^{-1/2}$ with fixed transfer factors and centroid precision.} 
  \label{fig:velocity_threshold}
  \vspace{-10.0pt}
\end{figure}

At $i=15^\circ$, combining Equation~\ref{eq:vpeak_mass} with the inferred $\sigma_{v,\rm fit}$ gives planet-mass detection thresholds of approximately $0.4\,\Mjup$ and $0.2\,\Mjup$ after seven and 28 hours of integration, respectively, for the simulated $2\,M_\odot$ host and a planet at $R_{\rm p}=240\,$au. If the empirical velocity normalisation is scaled as $v_{\rm K}\propto R^{-1/2}$ \citep[as in e.g.][]{rabago+2021}, the corresponding thresholds at $120\,$au become $0.27\,\Mjup$ and $0.14\,\Mjup$, while keeping fixed the effects of radiative transfer, beam convolution, and centroid precision measured at $240\,$au. For a self-similar $1\,M_\odot$ system, for which the absolute velocity perturbation scales as $M_*^{1/2}$, the corresponding planet-mass thresholds are $0.27\,\Mjup$ and $0.14\,\Mjup$ at $240\,$au, and $0.19\,\Mjup$ and $0.10\,\Mjup$ at $120\,$au, for seven and 28 hours, respectively.
At $i=30^\circ$, the seven-hour centroid precision is $2.8\,\msec$, compared with $2.3\,\msec$ at $15^\circ$. The stronger recovered signal gives $5\sigma$ planet mass thresholds of $0.30$ and $0.15\,\Mjup$ after seven and 28 hours, respectively, at 240\,au, despite the slightly poorer centroid precision. Fig.~\ref{fig:velocity_threshold} summarises these results. 
We also tested $i=45^\circ$, but the initial smooth-disc cube already exhibits coherent residuals with substantial backside contamination, with a spatial RMS of $\sim\!37\,\msec$ in the planet-wake sector, complicating separation of the planet-driven response from the background. 

A complementary rotated-aperture test in the seven-hour $i=15^\circ$ simulation gives a 95th-percentile peak-residual mass equivalent of $\sim\!0.56\,\Mjup$. This provides a structure reference for this particular simulation, but it is neither a $5\sigma$ threshold nor a universal detection floor because the control apertures are correlated and contain the planet's extended spiral response. Resolving this coherent large-scale response would nevertheless enable simultaneous modelling of the planet properties and hydrodynamical disc parameters \citep[e.g.][]{maio+2025}, providing independent constraints on both the embedded companion and its host disc. 

\paragraph{Dependence on source distance and planet orbital radius.}
Placing the source further away, from 101 to $150\,$pc, moderately decreases the centroid peaks by approximately $4\%$ and increases the seven-hour planet-mass thresholds by approximately $8\%$, from $0.38$ to $0.41\,\Mjup$ at $15^\circ$ and from $0.30$ to $0.32\,\Mjup$ at $30^\circ$, for the ALMA2040 Briggs $+0.5$ beam.
We also explored recovery at smaller orbital radii using planar \textsc{wakeflow} velocity injections \citep{hilder+2023}, retaining the fiducial planet-to-star mass ratio, seven-hour integration time and disc inclination of $15^\circ$, and applying an empirical velocity gain of three to approximate the beam-convolved planet-wake signal in the hydrodynamical reference described in Sect.~\ref{sec:planetflows}. At $150\,$pc, the adopted $0.095''\!\times\!0.065''$ beam has a major-axis FWHM of $\sim\!14\,$au. The peak planet-wake contrast relative to the smooth control is approximately 6 at eight beam widths and beyond. At smaller radii, it reaches 3.5 at four beam widths, decreases to $\sim\!2$ at three beam widths, and falls below unity at two beam widths. This suggests that targeted $\sim\!10\,$h ALMA2040 observations at $\sim\!0.05''$ resolution could bring kinematic signatures of $0.5$--$1.0\,\Mjup$ planets around $1$--$2\,M_\odot$ stars within reach down to $\sim30\,$au, provided sufficient line-brightness sensitivity and control of coherent disc-model residuals. 


\section{Synergies with current and future facilities}\label{sec:synergies}\vspace{-5.0pt}
The impact of ALMA2040 will be even broader thanks to synergies with forthcoming facilities, primarily the ELT. First, (sub)mm continuum and kinematics are complementary to ELT/METIS and MICADO high contrast imaging and ELT/METIS IR kinematics for the detection of (circumplanetary emission around) embedded planets, both in terms of contrast and separation. 
Secondly, (sub)mm kinematic planet detection, especially if associated with circumplanetary line emission, can provide constraints on the mass of embedded planets. Complementary detections in the IR will provide planet luminosities and thus the opportunity to build a planet~mass-luminosity diagram. This diagram is essential to test protoplanet evolutionary tracks~and discriminate between formation by disc instability and core accretion \citep[e.g.,][]{Spiegel&Burrows2012}. By the 2040's, two more decades of radial velocity monitoring, the forthcoming \textit{Gaia} data releases, and the Roman Space Telescope will have provided a complete characterisation of the mature gas giants within $30\,{\rm au}$ \citep[e.g.,][]{Wilson2023,Lammers&Winn2026}. This census will enable for the first time to statistically assess the role of migration in the formation of giant planets by comparing mature exoplanets with those inferred from continuum substructures. Observations with the ngVLA will also provide complementary information on CPD flux densities at lower frequencies: this synergy will be helpful to better estimate their masses.\vspace{-10.0pt}



\section*{Acknowledgements}\vspace{-5.0pt}
We thank the referees for their constructive feedback which substantially improved the quality of our work. This work was developed as part of the broader ALMA2040 community effort. We acknowledge the Lorentz Center and NOVA for supporting the community workshops and coordination activities that helped initiate and shape the ALMA2040 White Paper process.\\\vspace{-10.pt} 

\noindent\bibliographystyle{apj_mod}
\bibliography{bibliography}
\end{document}